\documentclass[runningheads]{llncs}
\usepackage{adjustbox}
\usepackage{blindtext}
\usepackage{rotating}
\usepackage[toc,page]{appendix}
\usepackage{verbatim}
\usepackage{marvosym}
\usepackage{syntax}
\usepackage{graphicx}
\usepackage{fullpage}
\usepackage[x11names]{xcolor}
\usepackage{soul}

\usepackage{amsthm}
\usepackage{amsmath}

\usepackage{amssymb, stmaryrd, mathpartir}

\usepackage{tikz}
\usetikzlibrary{cd}
\usepackage{hyperref}
\usepackage{tcolorbox}
\usepackage{amsfonts}

\usepackage[marginal]{footmisc}

\usepackage{algorithm,algorithmic}
\RequirePackage{algorithm}

\makeatletter
\newenvironment{breakablealgorithm}
  {
    \begin{center}
      \refstepcounter{algorithm}
      \hrule height.8pt depth0pt \kern2pt
      \parskip 0pt
      \renewcommand{\caption}[2][\relax]{
        {\raggedright\textbf{\fname@algorithm~\thealgorithm} ##2\par}%
        \ifx\relax##1\relax 
          \addcontentsline{loa}{algorithm}{\protect\numberline{\thealgorithm}##2}%
        \else 
          \addcontentsline{loa}{algorithm}{\protect\numberline{\thealgorithm}##1}%
        \fi
        \kern2pt\hrule\kern2pt
     }
  }
  {
     \kern2pt\hrule\relax
   \end{center}
  }

\usepackage{multirow} 

\usepackage[utf8]{inputenc}
\DeclareUnicodeCharacter{2200}{$\forall$}
\DeclareUnicodeCharacter{2203}{$\exists$}
\DeclareUnicodeCharacter{03BE}{$\xi$}
\DeclareUnicodeCharacter{0398}{$\theta$}
\DeclareUnicodeCharacter{2208}{$\in$}

\usepackage[english]{babel}

\usepackage{xcolor}
\usepackage{listings}
\definecolor{dkgreen}{rgb}{0,0.6,0}
\definecolor{ltblue}{rgb}{0,0.4,0.4}
\definecolor{dkviolet}{rgb}{0.3,0,0.5}
\definecolor{dkblue}{rgb}{0,0,.6}

\lstdefinelanguage{Coq}{ 
    mathescape=true,
    texcl=false, 
    morekeywords=[1]{Section, Module, End, Require, Import, Export,
        Variable, Variables, Parameter, Parameters, Axiom, Hypothesis,
        Hypotheses, Notation, Local, Tactic, Reserved, Scope, Open, Close,
        Bind, Delimit, Definition, Let, Ltac, Fixpoint, CoFixpoint, Add,
        Morphism, Relation, Implicit, Arguments, Unset, Contextual,
        Strict, Prenex, Implicits, Inductive, CoInductive, Record,
        Structure, Canonical, Coercion, Context, Class, Global, Instance,
        Program, Infix, Theorem, Lemma, Corollary, Proposition, Fact,
        Remark, Example, Proof, Goal, Save, Qed, Defined, Hint, Resolve,
        Rewrite, View, Search, Show, Print, Printing, All, Eval, Check,
        Projections, inside, outside, Def},
    morekeywords=[2]{forall, exists, exists2, fun, fix, cofix, struct,
        match, with, end, as, in, return, let, if, is, then, else, for, of,
        nosimpl, when},
    morekeywords=[3]{Type, Prop, Set, true, false, option},
    morekeywords=[4]{pose, set, move, case, elim, apply, clear, hnf,
        intro, intros, generalize, rename, pattern, after, destruct,
        induction, using, refine, inversion, injection, rewrite, congr,
        unlock, compute, ring, field, fourier, replace, fold, unfold,
        change, cutrewrite, simpl, have, suff, wlog, suffices, without,
        loss, nat_norm, assert, cut, trivial, revert, bool_congr, nat_congr,
        symmetry, transitivity, auto, split, left, right, autorewrite},
    morekeywords=[5]{by, done, exact, reflexivity, tauto, romega, omega,
        assumption, solve, contradiction, discriminate},
    morekeywords=[6]{do, last, first, try, idtac, repeat},
    morecomment=[s]{(*}{*)},
    showstringspaces=false,
    morestring=[b]",
    morestring=[d]’,
    tabsize=3,
    extendedchars=false,
    sensitive=true,
    breaklines=false,
    basicstyle=\small,
    captionpos=b,
    columns=[l]flexible,
    identifierstyle={\ttfamily\color{black}},
    keywordstyle=[1]{\ttfamily\color{dkviolet}},
    keywordstyle=[2]{\ttfamily\color{dkgreen}},
    keywordstyle=[3]{\ttfamily\color{ltblue}},
    keywordstyle=[4]{\ttfamily\color{dkblue}},
    keywordstyle=[5]{\ttfamily\color{dkred}},
    stringstyle=\ttfamily,
    commentstyle={\ttfamily\color{dkgreen}},
    literate=
    {\\forall}{{\color{dkgreen}{$\forall\;$}}}1
    {\\exists}{{$\exists\;$}}1
    {<-}{{$\leftarrow\;$}}1
    {=>}{{$\Rightarrow\;$}}1
    {==}{{\code{==}\;}}1
    {==>}{{\code{==>}\;}}1
    {->}{{$\rightarrow\;$}}1
    {<->}{{$\leftrightarrow\;$}}1
    {<==}{{$\leq\;$}}1
    {\#}{{$^\star$}}1 
    {\\o}{{$\circ\;$}}1 
    {\@}{{$\cdot$}}1 
    {\/\\}{{$\wedge\;$}}1
    {\\\/}{{$\vee\;$}}1
    {++}{{\code{++}}}1
    {~}{{$\sim$}}1
    {\@\@}{{$@$}}1
    {\\mapsto}{{$\mapsto\;$}}1
    {\\hline}{{\rule{\linewidth}{0.5pt}}}1
}[keywords,comments,strings]

\usepackage{multicol}

\begin{document}
\title{A Natural Formalized Proof Language}
%
%

\author{Lihan Xie \and Zhicheng Hui \and Qinxiang Cao\textsuperscript{(\Letter)}}
\authorrunning{L. Xie et al.}
%
\institute{Shanghai Jiao Tong University, Shanghai, China\\
\email{\{sheringham,laplace$\_$demon\}@sjtu.edu.cn, caoqinxiang@gmail.com}}
\maketitle              
\makeatletter\def\Hy@Warning#1{}\makeatother
\footnotetext{L. Xie and Z. Hui contributed equally to this work.}
\begin{abstract}
Artificial intelligence assisted mathematical proof has become a highly focused area nowadays. One key problem in this field is to generate formal mathematical proofs from natural language proofs. Due to historical reasons, the formal proof languages adopted by traditional theorem provers were not intended to represent natural language proofs. Therefore, they are not well-suited for the aforementioned tasks and proof-checking work for educational purposes. In this paper, we design a proof language and its corresponding abstract syntax tree and implement a proof checking tool for it. This language can be easily converted from natural language, thus providing a rich corpus of formal proof. Additionally, it supports the handling of issues in informal proofs through static analysis, and enhances the expressive power of the language by introducing the structure of partial proofs. This design combines the expressiveness of natural language and the accuracy of formal language, resulting in an improved mathematical proof language.
\end{abstract}

\section{Introduction} \label{intro}
Formal mathematical proofs are based on rigorous reasoning in formal logic, providing a completely accurate proof process that can be automatically verified by computers. Formalizing informal proofs can make them more convincing, as seen in the formal proof of the Four-Color Theorem\cite{appel1989every,gonthier2008formal}. Additionally, automated theorem proving relies on formal proof languages to generate rigorous proofs. The rise of large language models in recent years has spurred research in AI-assisted mathematical proof, particularly in the fields of automated theorem proving and transforming informal proofs into formal proofs\cite{lample2022hypertree,wu2022autoformalization}. This means that large language models can directly generate formal proofs or indirectly transform natural language proofs into formal proofs. These formal proofs can then be verified by theorem provers, ensuring their correctness.

Due to historical reasons, early versions of theorem provers were primarily focused on ensuring the correctness of proofs, rather than directly modelling natural language proofs. Subsequent developments, such as the Ltac\cite{David2000Ltac} tactic language in Coq\cite{barras1999coq} and Isar\cite{Wenzel1999IsarA} proof language in Isabelle\cite{paulson1994isabelle}, aimed to help with proof constructions. Commands in these languages can be seen as transformations that modify the current proof goal, where each proof goal comprises named premises and the conclusion that needs to be proved. However, it is still difficult to translate natural language proofs word-for-word into such formal proof languages. Consequently, for current applications like AI-assisted mathematical proofs or automated proof grading for educational purposes, these formal languages seem to be insufficient.

In the following example shown in Fig.\ref{ex1} about the proof of the monotone convergence theorem through the supremum theorem, we can observe several characteristics of natural language proofs and the difficulties consequent upon the task of formalization using existing formal proof languages like Coq. We will also demonstrate how our work circumvents these difficulties.

\begin{figure}

\begin{quotation}
\noindent Monotone Convergence Theorem:
For every sequence of real numbers $(a_n)_{n \in \mathbb{N}}$, $(a_n)_{n \in \mathbb{N}}$ converges if it is monotonically increasing and bounded above.

\begin{proof}

(1) Assume $(a_n)_{n \in \mathbb{N}}$ is monotonically increasing and bounded above. 

(2) By supremum theorem, there exists $A$ such that $A = \sup \{a_n\}$.

(3) We use the definition of limit to show that $\lim\limits_{n \to +\infty}{a_n}=A$.

(4) For every $\varepsilon > 0$, by the definition of least upper bound, there exists an integer $N$ such that $a_N > A - \varepsilon$.

(5) Since $\{a_n\}$ increases, for every $n$, $n > N$ implies $a_n > a_N$.

(6) Since $A$ is an upper bound of  $(a_n)_{n \in \mathbb{N}}$, for every $n$, $n > N$ implies $a_n < A$.

(7) Consequently, for every $n$, $n > N$ implies $A - \varepsilon < a_n < A + \varepsilon$.

(8) By the definition of convergence, we have $\lim\limits_{n \to +\infty}{a_n}=A$,

(9) which proves the theorem.
\end{proof}
\end{quotation}
\caption{Proof of monotone convergence theorem.} \label{ex1}
\end{figure}

\newpage

\begin{description}

\item[Partial Transformation of Proof Goal.] The concept of proof goal models the task of a mathematical proof, i.e. deriving the conclusion from premises and proven results. Following the steps of a proof, we implicitly transform the proof goal by introducing new variables, proving intermediate results, or posing subgoals. For example, we pose a subgoal in line 3, thereby transforming the proof goal into two subsequent proof goals: (1) proving $\lim\limits_{n \to +\infty}{a_n}=A$, (2) and proving the conclusion with the intermediate result $\lim\limits_{n \to +\infty}{a_n}=A$ proven. However, there are circumstances where the subsequent proof goal cannot be found directly: We pose an assumption $\epsilon > 0$ in line 4 and derive the result $\forall n > N, A - \epsilon < a_n < A + \epsilon$ in line 7 under the assumption. The overall process proves the result $\forall \epsilon > 0, \exists N \in \mathbf{N}, \forall n > N, A - \epsilon < a_n < A + \epsilon$, which is not explicitly stated in line 4. Consequently, the transformation of the proof goal in line 4 is partial. Since each tactic in tactic languages should represent a clearly defined, complete transformation of the current proof goal, formalizing the proof using tactic languages would involve the extra task of determining the subsequent proof goal. For example, by inferring the proposition to be filled in a Coq ``assert'' tactic. \textit{In our work, a structure of \hyperref[pp]{partial proof} is included in the design of proof language to model this pattern of proof, it eliminates the need for additional inferences or structural modifications on the proof during the task of formalization.}

\item[Context-dependent Semantics.] Depending on the context, the same natural language statement could have multiple interpretations, of which the semantic differences may be subtle. For example, when we write ``there exists'' in the proof, at least three interpretations are possible. (1) In the context of proving an existential statement, a satisfying value has been found for one of the existential quantifiers in the conclusion to be proved, (2) or a proposition beginning with an existential quantifier is stated, (3) or similar to the second case, except that the qualified variable becomes a free variable and can be used later, as in line 2 and line 4 of the example, where variables $A$ and $N$ are used in line 3 and line 5, respectively. In order to use the tactic language of Coq, all those semantic variances must be explicitly formulated. Namely by an ``exists'' tactic for the case (1), an existentially quantified variable in the proposition for the case (2) and a free variable in the proposition for the case (3). Determining the correct semantic interpretation would require analysing the context during the process of formalization. Not only is such an hidden task of analysis generally harder to perform on an unstructured natural language proof, but it is also indirect on a tactic proof, as tactics do not explicitly contain information on proof goals. That explains the difficulties faced by the tactic languages as object languages of automatic formalization. \textit{In our work, the proof language is designed to resemble natural language in order to streamline the formalization. Moreover, the resemblance allows the proof language to temporarily preserve the context-dependent semantics, thus allowing the resolution to be postponed until we can perform \hyperref[4.3]{static analysis} on the formalized version of the proof.}

\item[Overloading of Notation.] Sometimes a notational convention may be employed though not being mathematically rigorous. For example, the appearance of $\{a_n\}$ in line 2 and line 5 does not represent a singleton but rather the set containing all elements of the sequence $a$. Another typical example is the extensive usage of $f(x)$ for representing the function $f$ itself. Similar to context-dependent semantics, all mathematical formulas should be written rigorously in Coq. \textit{By performing \hyperref[4.3]{static analysis} on the entire formal proof, the precise meaning of each expression can also be inferred.}

\end{description}

In summary, to address the problems mentioned above, we design a natural-language-like formal proof language for modeling mathematical proofs. We add partial proof structures to make them more similar to the natural language proofs. Furthermore, after transforming the natural formal proof into an abstract syntax tree by a parser, we can perform static analysis on it to resolve context-dependent semantics and overloading of notation.

Accordingly, we implement a framework called ProofGrader for checking mathematical proofs automatically. For each step of reasoning, corresponding solvers are chosen heuristically in order to check its correctness. Apart from the proof checking kernel, the rest of the system is designed to be highly modular so that our system may fit into different usage scenarios: It is possible to alter the mathematical objects and the usable theorems involved in the proof, the acceptable forms of proof steps, or the solvers used for checking. For example, to serve educational purposes, mild solvers fitting the level of human intuition can be plugged in instead of powerful ones, and advanced theorems can be temporarily disabled until they are proved or taught later.

In the rest of this paper, we will first present the design of our proof language in Section \ref{2}, with elaborations on several important elements. After that, we give the formal semantics of our proof language in Section \ref{3}. We then show the schema of the workflow of our checker in Section \ref{4}. A detailed description of our solvers is developed in Section \ref{5}. Section \ref{evaluation} gives an evaluation of the proof checker. And then we will introduce some related works in Section \ref{related}. Finally, Section \ref{Conclusion} is devoted to a conclusion.

\section{Proof Language Design} \label{2}

We focus on defining a natural-language-like proof language, whose structure faithfully reflects that of natural language. Thus, our natural formal proof language is provided with the hierarchy of natural language proof. At the top level are proof steps, what follows are propositions and terms. A certain amount of investigations of natural language proofs are done to incorporate common proof patterns within our proof language. Such a proof language follows predefined grammar rules but can be read directly as natural language, an explanation of its grammar can be found in Appendix \ref{AA}.

To perform proof checking, the natural formal proof language is further translated into an abstract syntax tree via a \hyperref[4.2]{parser}. Below shows a subset of its definitions. The definitions related to \texttt{term} and \texttt{prop} can be found in the Appendix \ref{BB}.

\begin{multicols}{2}
\begin{grammar}

<proof> ::= `ProofAction' <action> <proof>
\alt `PoseWithoutProof' <fwd> <prop> <proof>
\alt `PoseAndProve' <fwd> <prop> <proof> <proof>
\alt `ClaimSuffice' <bwd> <prop> <proof>
\alt `ProveSuffice' <bwd> <prop> <proof> <proof>
\alt `ConclWithoutProof' <fwd>
\alt `ConclAndProve' <fwd> <proof>
\alt `PosePartialProof' <poseAction> <proof> <proof>
\alt `EndPartialProof'

<fwd> ::= `FNoHint'
\alt `FDefinition' <definition>
\alt `FTheorem' <theorem>
\alt `FAddEqn' \{\textit{identifier}\}$^*$
\alt `FDeriBothTerms' <identifier>
\alt ...

\columnbreak

<bwd> ::= `BNoHint'
\alt `BContra'
\alt ...

<action> ::= `AIntros' <identifier>
\alt `AExists' <term>
\alt `ASuppose' <prop>
\alt `ASet' <identifier> <term>
\alt `ASetProp' <prop>
\alt `AExistVar' <identifier>

<poseAction> ::= `APoseVar' <identifier> \{\textit{prop}\}$^*$
\alt `APoseProp' <prop>

\end{grammar}
\end{multicols}

\noindent In the following, we will explain the meaning of each proof structure. We will refer to the names of abstract syntax tree nodes because these names correspond to components of the proof language. Some of our proof structures have similar functionalities to those of certain tactic language, while others have no counterpart, which enables us to better express natural language proof.

\paragraph{ProofAction} The field \texttt{action} constitutes an operation on the proof goal, and the field \texttt{proof} refers to subsequent proof steps.

The design of \texttt{ProofAction} refers to some tactic languages, such as the tactic \textit{intro} and \textit{exists} of Coq\cite{10.5555/1965123}, which deal with the quantifiers in the proposition to be proved. That corresponds to our proof action \texttt{AIntros} and \texttt{AExists}. However, the variety of proof actions is richer than that of tactic language. For example, \texttt{ASuppose} introduces a premise in the conclusion rather than a variable, which is also expressed by the tactic \textit{intro} in Coq. Such a difference is due to the fact that the proof language models the natural language directly, and thus is more in line with human intuition.

\texttt{ASet} binds a name to a term, to which the proof can refer later, and the existence of the term will be verified by the checker to ensure mathematical rigor.

\begin{figure}
\centering
\includegraphics[width=15.5cm]{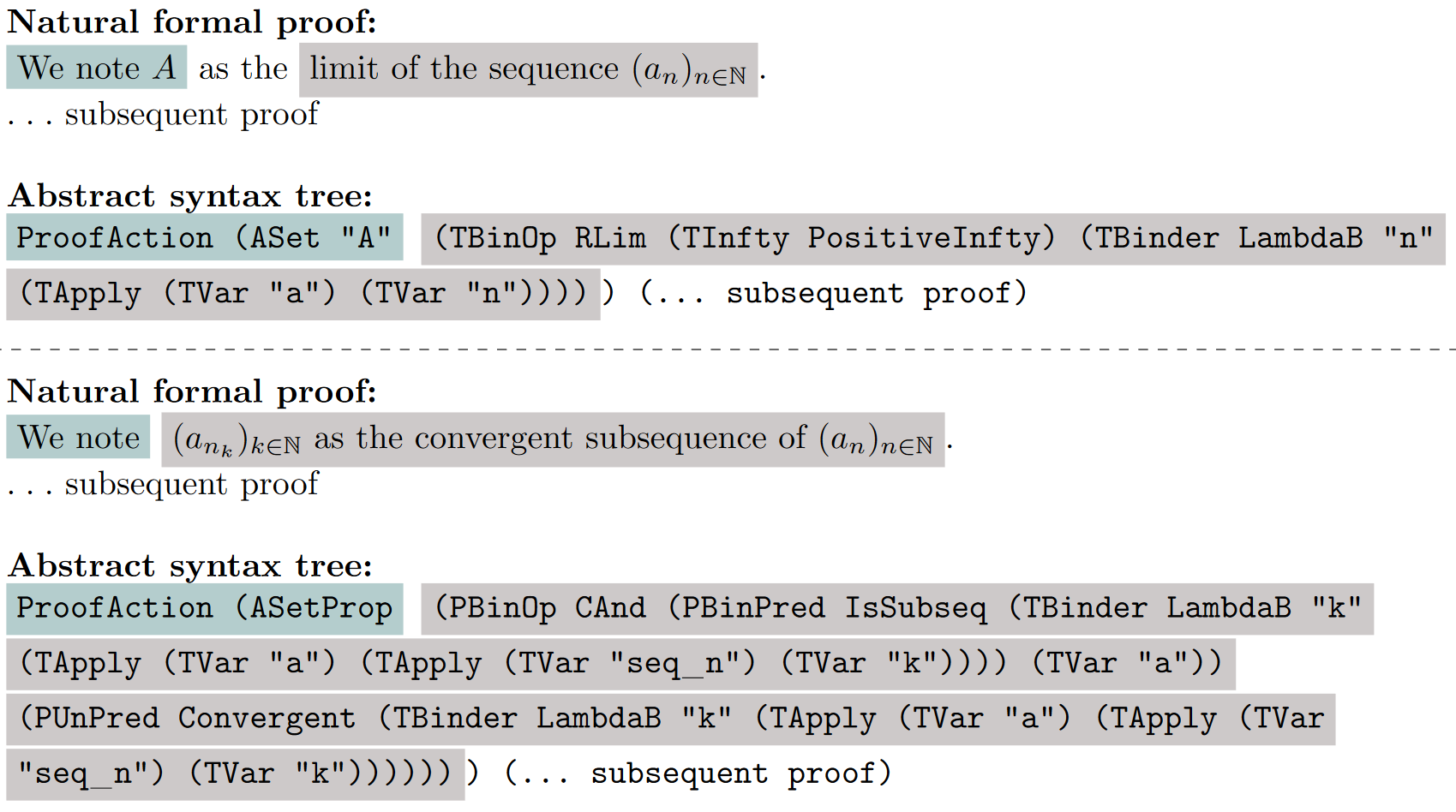} 
\caption{Example of ProofAction.} \label{ex2}
\end{figure}

\texttt{ASetProp} is a generalization of \texttt{ASet}. Instead of introducing a new variable through an equation, \texttt{ASetProp} introduces new variables through a proposition. Fig.\ref{ex2} below contrasts the usage of these two proof actions, the one-to-one correspondence between natural language and our proof language is marked in background color.

Finally, \texttt{AExistVar} indicates the action of instantiating the variable mentioned by an existential quantifier in the last premise in the proof goal. The usage of \texttt{AExistVar} is further demonstrated in Section \ref{4.3} when we perform static analysis.












\noindent In the upper part of Fig\ref{ex2}, the variable $A$ is set equal to the limit of a sequence, the checker then verifies whether the limit exists. In the lower part, a variable $k$ is introduced implicitly due to the introduction of a subsequence, the checker then verifies whether $(a_n)_{n \in \mathbb{N}}$ admits a convergent subsequence. Both of the two proof steps will add corresponding assumptions to the proof goal.

\paragraph{PoseWithoutProof \& PoseAndProve} These two components correspond to forward reasoning. The field \texttt{fwd} indicates the method involved in forward reasoning. The field \texttt{prop} denotes the proposition that the step proves. The last field \texttt{proof} still refers to subsequent proof steps.

\begin{figure} 
\centering
\includegraphics[width=15.6cm]{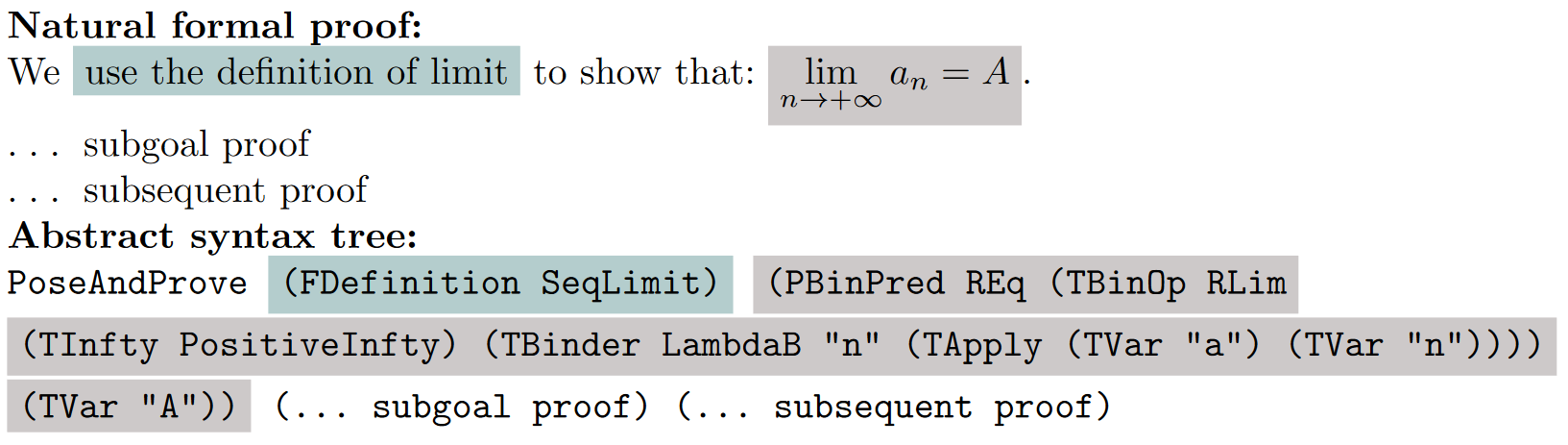} 
\caption{Example of PoseAndProve.} \label{ex3}
\end{figure}

Depending on the complexity of the reasoning, one may choose to provide a proof as a justification of the reasoning or just let the checker figure it out. This makes the difference of \texttt{PoseWithoutProof} and \texttt{PoseAndProve}. \texttt{PoseAndProve} carries an extra field \texttt{prop}, which is a complete proof showing how to derive its result. This design is also widespread in tactic languages. An example is the tactic \textit{assert} in Coq, which poses a subgoal to be proved. Fig.\ref{ex3} shows its usage scenario, which corresponds to the line 3 of Fig.\ref{ex1}.

\noindent Since the hint of using the definition of limit is far from sufficient to prove the result, an additional proof is provided to complete this task. The proven subgoal is then available in the subsequent proof.

The set of possible methods \texttt{fwd} is rather rich. To name a few, \texttt{FTheorem} denotes applying a theorem, \texttt{FAddEqn} denotes adding several equations together to get a new one, and \texttt{FDeriBothTerms} denotes taking a derivative from both sides of an equation. If no method is indicated, \texttt{FNoHint} is filled in. It is quite easy to incorporate new methods so this set is highly expandable.

\paragraph{ClaimSuffice \& ProveSuffice} These two components correspond to backward reasoning. The first field \texttt{bwd} indicates the method involved in backward reasoning, the second field \texttt{prop} denotes the proposition the step proves and the last field \texttt{proof} refers to subsequent proofs. The distinction between \texttt{ClaimSuffice} and \texttt{ProveSuffice} follows that between \texttt{PoseWithoutProof} and \texttt{PoseAndProve}.

Backward reasoning signifies that they start from the conclusion to be proved. The result provided is supposed to imply goal, which will then become the new goal.

\paragraph{ConclWithoutProof \& ConclAndProve} These two components correspond to the step of deriving the conclusion and terminating the proof, only a field \texttt{fwd} is presented to indicate the method involved. \texttt{ConclAndProve} allows the choice of presenting an extra explanatory proof when the derivation of the conclusion is not immediate.

Basically, \texttt{ConclWithoutProof} corresponds to the last step of the proof, such as the line 9 ``which proves the theorem'' in Fig.\ref{ex1}, and does not carry much information. But such a concluding remark imitates natural language proof, and the similarity to natural language characterizes our proof language.

\paragraph{PosePartialProof \& EndPartialProof} \label{pp}

In theorem provers, we always need to explicitly keep a record of the current proof goal. In most cases, this is the overall proposition to be proved. We can also pose a subgoal and prove it subsequently, a process represented by the \texttt{PoseAndProve} component in our proof language.

\begin{figure}
\centering
\includegraphics[width=12cm]{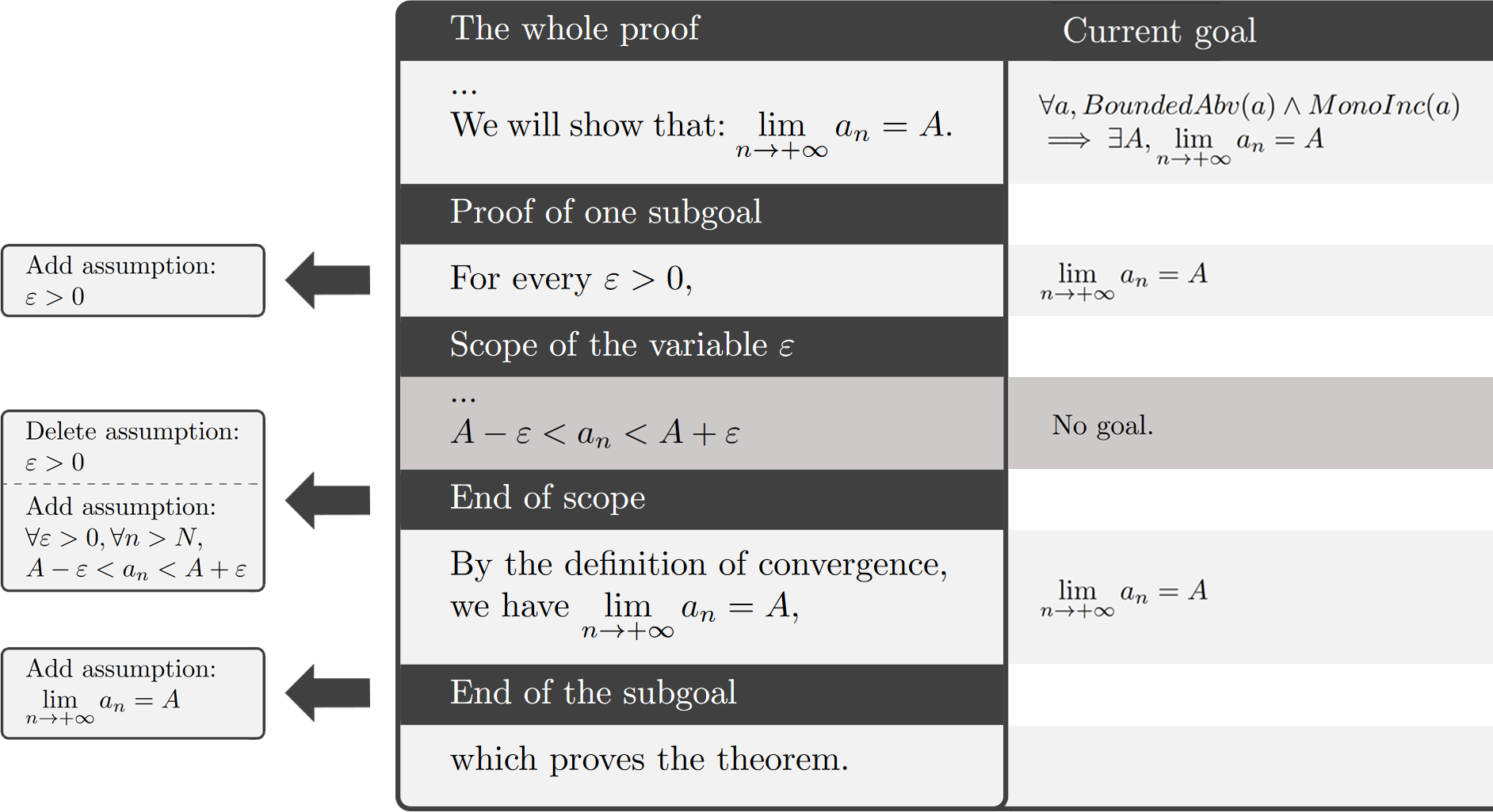}
\caption{Illustration of how the proof goal changes.} \label{fig1}
\end{figure}

As discussed in Section \ref{intro}, natural language proofs may also involve partial transformations of the proof goal, in the sense that subgoals are not explicitly stated in advance. Instead, We simply pose certain assumptions and proceed to derive the subgoals. Refer to Fig.\ref{fig1} for an illustration of how the different forms of proof goal are in the example Fig.\ref{ex1}. The dark gray part of the proof denotes what we term as a ``partial proof'', aligning with the \texttt{PosePartialProof} and \texttt{EndPartialProof} components in our proof language.

The notion of partial proof can be considered as one innovation of our proof language. Compared with tactic languages, it offers us extra convenience in modelling natural language proofs.

\noindent A partial proof starts with a list of variables and propositions, which then become the temporary assumptions. This information is carried by the first field \texttt{poseAction} is the grammar. \texttt{APoseVar} poses a new temporary variable along with related assumptions, while \texttt{APoseProp} poses a proposition. The former field \texttt{proof} carries the partial proof, which shall be terminated by \texttt{EndPartialProof}. After the termination of partial proof, all proven propositions will be altered to its proper form — without free variables. These altered propositions will become henceforth available in the subsequent proof — the latter field \texttt{proof}.

\begin{figure}
\centering
\includegraphics[width=15.5cm]{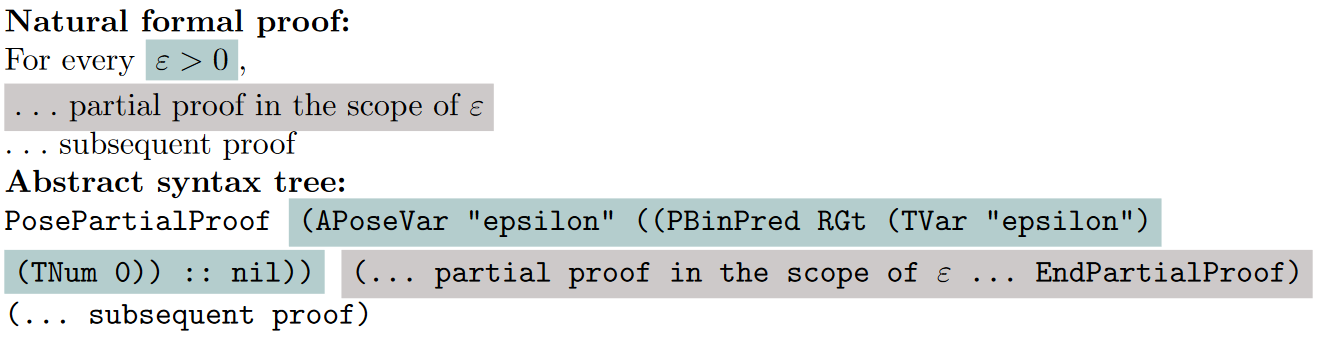}
\caption{Example of PosePartialProof.} \label{ex4}
\end{figure}

In Fig.\ref{fig1}, the variable $\varepsilon$ and the proposition $\varepsilon > 0$ are posed to start a partial proof. When reaching the result $\forall n > N, A - \varepsilon < a_n < A + \varepsilon$ at the end of the partial proof, it is expanded to $\forall \varepsilon>0, \forall n > N, A - \varepsilon < a_n < A + \varepsilon$, which is subsequently used to prove convergence. Fig.\ref{ex4} shows the how a partial proof looks like, which corresponds to line 4 of Fig.\ref{ex1}.

\section{The Formal Semantics of Proof Language} \label{3}

In this section, we demonstrate a selected set of the formal semantics of our proof language. Since each proof step defines a transformation of the proof goal, we use the notation $(pr, pg) \rightarrow (pr', pg')$ to denote that the proof goal $pg'$ results from the proof goal $pg$ after one or more proof steps in $pr$, with $pr'$ continuing the proof. A proof goal is defined as a pair $(A, C)$, where $A$ is a list of premises and $C$ a conclusion that needs to be proved. In the case of partial proof, the conclusion $C$ does not exist and we note it as $\square$.

For convenience, we represent a list of premises as a set of propositions, and we write the triple $(pr, A, C)$ to represent the pair $(pr, (A, C))$. We use the notation $FV(A)$ to denote the set of free variables in the set of propositions $A$, the same notation $FV(t)$ is also used to denote the set of free variables in the term $t$. We also define a constant $QED$ such that the proof is successfully completed when $(pr, A, C) = QED$.

\paragraph{Proof Action}

\[
\inferrule*[Right=Intros]
  { v \notin FV(A \cup \{\forall x.C\}) }
  { (\textbf{ProofAction} \; (\textbf{AIntros} \; v) \; pr, A, \forall x. C) \rightarrow (pr, A, C[v/x]) }
\]
\vspace{-4mm}
\[
\inferrule*[Right=Exists]
  { FV(t) \subset FV(A \cup \{\exists x.C\}) }
  { (\textbf{ProofAction} \; (\textbf{AExists} \; t) \; pr, A, \exists x. C) \rightarrow (pr, A, C[t/x]) }
\]
\vspace{-4mm}
\[
\inferrule*[Right=Suppose]
  {\text{PropEquivalent} \; P \; Q}
  { (\textbf{ProofAction} \; (\textbf{ASuppose} \; P) \; pr, A, Q \Rightarrow C) \rightarrow (pr, A \cup \{P\}, C) }
\]
\vspace{-4mm}
\[
\inferrule*[Right=Set]
  {\text{TermWellDefined} \; t \; A \; \; \; \; \; v \notin FV(A \cup \{C\}) }  { (\textbf{ProofAction} \; (\textbf{ASet} \; v \; t) \; pr, A, C) \rightarrow (pr, A \cup \{v=t\}, C) }
\]
\vspace{-4mm}
\[
\inferrule*[Right=SetProp]
  {\text{PropWellDefined} \; P \; A}
  { (\textbf{ProofAction} \; (\textbf{ASetProp} \; P) \; pr, A, C) \rightarrow (pr, A \cup \{P\}, C) }
\]
\vspace{-4mm}
\[
\inferrule*[Right=ExistVar]
  {v \notin FV(A \cup \{\exists x.P\} \cup \{C\})}
  { (\textbf{ProofAction} \; (\textbf{AExistVar} \; v) \; pr, A \cup \{\exists x. P\}, C) \rightarrow (pr, A \cup \{\exists x. P\} \cup \{P[v/x]\}, C) }
\]

\noindent Proof actions are proof steps that directly manipulates the proof goal, usually when some conditions are met. In the semantic definitions above, $\text{PropEquivalent} \; P \; Q$ means the proposition $P$ and $Q$ are equivalent, $\text{TermWellDefined} \; t \; A$ means the term $t$ is well-defined under the set of premises $A$, and $\text{PropWellDefined} \; P \; A$ means the proposition $P$ is well-defined and satisfied under the set of premises $A$. These predicates are implemented as functions within our proof checking system, however, they are too cumbersome to define formally.

The behavior of $\textbf{AIntros} \; v$ is to remove the universal quantifier in the conclusion $\forall x.C$ and replace the variable $x$ in $C$ by the variable $v$, when the variable $v$ does not occur freely in the proof goal. Similarly, the behavior of $\textbf{AExists} \; t$ is to remove the existential quantifier in the conclusion $\exists x.C$ and replace the variable $x$ in $C$ by the term $t$, when the term $t$ does not contain free variables other than those in the proof goal. And the behavior of $\textbf{ASuppose} \; P$ is to remove the implication in the conclusion $Q \Rightarrow C$ and add $P$ to the premises when $P$ and $Q$ are equivalent propositions.

$\textbf{ASet} \; v \; t$ adds the equation $v = t$ to the premises, where $v$ does not occur freely and the term $t$ must be well-defined in the proof goal. The requirement for well-definiteness ensures that a term does not appear before its existence has been proved in the proof goal. The behavior of $\textbf{ASetProp} \; P$ is similar, it incorporates the proposition $P$ possibly containing new variables into the premises. This can be useful when $P$ is not an equation.

Finally, when the last premise in the proof goal is an existential proposition, the statement $\textbf{AExistVar} \; v$ instantiates the existential quantifier with the variable $v$, by adding the properties verified by $v$ to the premises. It is required that the variable $v$ does not occur freely in the proof goal.

\paragraph{Direct Forward Proof}

\[
\inferrule*[Right=NoHint]
  {\text{PropDeducible} \; P \; A}
  { (\textbf{PoseWithoutProof} \:\: \textbf{FNoHint} \; P \; pr, A, C) \rightarrow (pr, A \cup \{P\}, C) }
\]
\vspace{-4mm}
\[
\inferrule*[Right=ApplyTheorem]
  {\text{TheoremApplicable} \; thm \; P \; A }
  { (\textbf{PoseWithoutProof} \; (\textbf{FTheorem} \; thm) \; P \; pr, A, C) \rightarrow (pr, A \cup \{P\}, C) }
\]
\vspace{-4mm}
\[
\inferrule*[Right=Concl]
  {\;}
  { (\textbf{ConclWithoutProof} \:\: \textbf{FNoHint}, A \cup \{C\}, C) \rightarrow QED }
\]
\vspace{-4mm}
\[
\inferrule*[Right=ConclWithTheorem]
  { \text{TheoremApplicable} \; thm \; C \; A }
  { (\textbf{ConclWithoutProof} \; (\textbf{FTheorem} \; thm), A, C) \rightarrow QED }
\]

\noindent The predicate $\text{PropDeducible} \; P \; A$ indicates that the proposition $P$ can be deduced from the set of premises $A$. Similarly, $\text{TheoremApplicable} \; thm \; P \; A$ signifies that the theorem $thm$ can be applied to the set of premises $A$ to yield the result $P$. It is important to note that the notion of deducibility here is algorithmic rather than theoretical: a proposition is considered deducible if it can be derived using our predefined set of solvers within a specified number of steps. The applicability of theorem is also defined by the theorem checker in our implementation. Again, we do not provide formal definitions but give descriptions in the following.

The behavior of $\textbf{PoseWithoutProof} \:\: \textbf{FNoHint}$ depends largely on the solvers. In the process of proof checking, the checker will send the proposition $P$ and the current proof goal $(A, C)$ to the solver manager, which checks whether the proposition $P$ can be derived. Then, the checker adds $P$ to the premises of the proof goal, and checks the rest of the proof $pr$.

Similarly, the behavior of $\textbf{PoseWithoutProof} \; (\textbf{FTheorem} \; thm)$ depends on the theorem checker in our implementation. The theorem checker checks two aspects: (1) whether the prerequisites for applying the theorem $thm$ are satisfied, (2) and whether $P$ can be derived as a conclusion from the theorem under those prerequisites.

The working process of the theorem checker begins with a pattern match between $P$ and the conclusion of the theorem $thm$. This pattern match instantiates the variables and identifies the prerequisites of the theorem. Then, the theorem checker searches in the proof goal to see if all the prerequisites are satisfied. If the theorem checker is unable to match the conclusion or find all the prerequisites, then this step of transformation cannot be realized.

$\textbf{ConclWithoutProof} \:\: \textbf{FNoHint}$ transforms the proof goal into $QED$ when the conclusion appears in the premises. $\textbf{ConclWithoutProof} \; (\textbf{FTheorem} \; thm)$ does the same transformation when the conclusion follows from the theorem $thm$ by the theorem checker.

The Table \ref{tab:table1} shows how the proof goal gets transformed when reaching line 2 of Fig.\ref{ex1}, where we apply the supremum theorem.

\begin{table}[!ht]
\begin{center}
\begin{tabular}{l|c}
\hline
\multicolumn{1}{c|}{\multirow{2}{*}{\textbf{Premises}}}                                                            & \multirow{2}{*}{\textbf{Conclusion}}         \\
\multicolumn{1}{c|}{}                                                                                     &                                     \\ \hline
\multirow{3}{*}{\begin{tabular}[c]{@{}l@{}} $\{ a_n\}$ has an upper bound\\ $\{ a_n\}$ is monotonically increasing\end{tabular}} & \multirow{5}{*}{there exists $A$ such that $A = \lim\limits_{n \to +\infty}{a_n}$} \\
                                                                                                          &                                     \\
                                                                                                          &                                     \\ \cline{1-1}
\multirow{2}{*}{\textbf{+} there exists $A$ such that $A = \sup \{a_n\}$}                                                                     &                                     \\
                                                                                                          &                                     \\ \hline
\end{tabular}
\end{center}
\caption{Transformation of proof goal when reaching line 2 of the proof in Fig.\ref{ex1}. The proposition marked with \textbf{+} is added by the transformation of line 2.}
\label{tab:table1}
\end{table}

\paragraph{Subgoal Forward Proof}

\[
\inferrule*[Right=Subgoal]
  { (pr, A, P) \rightarrow QED }
  { (\textbf{PoseAndProve} \; \textbf{FNoHint} \; P \; pr \; pr', A, C) \rightarrow (pr', A \cup \{P\}, C) }
\]

\noindent The subgoal proof statement \noindent $\textbf{PoseAndProve} \:\: \textbf{FNoHint}$ adds the proposition $P$ to the premises on the condition that the proof of the subgoal is successfully completed. In the process of proof checking, the checker first checks the subgoal proof, and then returns to the main proof with the proven result.

The Table \ref{tab:subgoal} shows how the proof goal gets transformed when reaching line 3 of Fig.\ref{ex1}, where the lines 3-8 constitutes a typical subgoal proof. It generates a new proof goal and provides a proof for it. The subgoal gets added to the premises as the newly added proof succeeds.

\begin{table}[!ht]
\begin{center}
\begin{tabular}{l|c}
\hline
\multicolumn{1}{c|}{\multirow{2}{*}{\textbf{Premises}}}                                                                                                                                                       & \multirow{2}{*}{\textbf{Conclusion}}                                                                            \\
\multicolumn{1}{c|}{}                                                                                                                                                                                &                                                                                                        \\ \hline
\multirow{5}{*}{\begin{tabular}[c]{@{}l@{}}$\{ a_n\}$ has an upper bound\\ $\{ a_n\}$ is monotonically increasing\\ there exists $A$ such that $A = \sup \{a_n\}$\\ $A = \sup \{a_n\}$\end{tabular}} & \multicolumn{1}{l}{\multirow{3}{*}{there exists $A$ such that $A = \lim\limits_{n \to +\infty}{a_n}$}} \\
                                                                                                                                                                                                     & \multicolumn{1}{l}{}                                                                                   \\
                                                                                                                                                                                                     & \multicolumn{1}{l}{}                                                                                   \\ \cline{2-2} 
                                                                                                                                                                                                     & \multirow{2}{*}{$A = \lim\limits_{n \to +\infty}{a_n}$}                                                \\
                                                                                                                                                                                                     &                                                                                                        \\ \hline
\end{tabular}
\end{center}
\caption{Transformation of proof goal when reaching line 3 of the proof in Fig.\ref{ex1}. The premises remain unchanged, but a new conclusion is generated.}
\label{tab:subgoal}
\end{table}

\paragraph{Partial Proof}  
{
\[   
\inferrule*[Right=\scriptsize PoseVar]
   {\scriptsize(pr, A \cup sP, \square) \rightarrow (\textbf{EndPartialProof}, A', \square) \; \; \; \; \;  \scriptsize FV(sP) \subset FV(A \cup \{C\}) \cup \{s\} }
  { \scriptsize (\textbf{PosePartialProof} \; (\textbf{APoseVar} \; s \; sP) \; pr \; pr', A, C) \rightarrow (pr', A \cup (\text{AddVarDep} \; (A' \setminus (A \cup lP)) \; s \; sP), C) } 
\]
\vspace{-4mm}
\[    
\inferrule*[Right=\scriptsize PoseProp]
  {\scriptsize (pr, A \cup \{P\}, \square) \rightarrow (\textbf{EndPartialProof}, A', \square) \; \; \; \; \; \scriptsize FV(\{P\}) \subset FV(A \cup \{C\}) }
  { \scriptsize (\textbf{PosePartialProof} \; (\textbf{APoseProp} \; P) \; pr \; pr', A, C) \rightarrow (pr', A \cup (\text{AddPropDep} \; (A' \setminus (A \cup \{P\})) \; P), C) }
\]
}
\noindent A partial proof first makes one or more assumptions and then deduces a series of results. After the partial proof terminates, these results are added with dependencies on the assumptions, by prefixing them with universal quantifiers and prerequisite conditions. The assumptions can be either (1) posing a new variable satisfying certain conditions, (2) or posing a hypothesis on the existing variables of the proof goal. These two cases correspond to the constructs $\textbf{APoseVar} \; s \; sP$ and $\textbf{APoseProp} \; P$ defined above, where $s$ refers to the name of the posed variable, $sP$ a set of assumptions on the posed variable and $P$ an assumption on existing variables. The two operators $\text{AddVarDep}$ and $\text{AddPropDep}$ add dependencies on the assumptions to the results derived during the partial proof. Similar to the subgoal proof, a $\textbf{PosePartialProof}$ statement causes the checker to first check the partial proof. After reaching $\textbf{EndPartialProof}$, which marks the end of the partial proof, the checker integrates the proof goal back to the main proof by modifying all the derived premises.

The Table \ref{tab:partial} shows how the proof goal gets transformed between lines 4-7 in Fig.\ref{ex1}, which represents a partial proof structure. It first makes an assumption $\epsilon > 0$ and then deduces several results. After that, these results are added with a prefix $\forall \epsilon > 0$ to indicate their dependency on the assumption $\epsilon > 0$.

\begin{table}[ht!]
\begin{center}
\begin{tabular}{l|c}
\hline
\multicolumn{1}{c|}{\multirow{2}{*}{\textbf{Premises}}}                                                                                                                                                                                                                                                                                                                                                                                                                                                                   & \multirow{2}{*}{\textbf{Conclusion}}                             \\
\multicolumn{1}{c|}{}                                                                                                                                                                                                                                                                                                                                                                                                                                                                                            &                                                         \\ \hline
\multirow{5}{*}{\begin{tabular}[c]{@{}l@{}}$\{ a_n\}$ has an upper bound\\ $\{ a_n\}$ is monotonically increasing\\ there exists $A$ such that $A = \sup \{a_n\}$\\ $A = \sup \{a_n\}$ \end{tabular}}                                                                                                                                                                                                                                                                                             & \multirow{5}{*}{$A = \lim\limits_{n \to +\infty}{a_n}$} \\
                                                                                                                                                                                                                                                                                                                                                                                                                                                                                                                 &                                                         \\
                                                                                                                                                                                                                                                                                                                                                                                                                                                                                                                 &                                                         \\
                                                                                                                                                                                                                                                                                                                                                                                                                                                                                                                 &                                                         \\
                                                                                                                                                                                                                                                                                                                                                                                                                                                                                                                 &                                                         \\ \hline
\multirow{6}{*}{\begin{tabular}[c]{@{}l@{}}\textbf{+} $\epsilon > 0$ \textbf{(partial proof start)}\\ \textbf{+} there exists $N$ such that $a_N > A - \epsilon$\\ \textbf{+} for all $n$, if $n > N$ then $a_n > a_N$\\ \textbf{+} for all $n$, if $n > N$ then $a_n < A$\\ \textbf{+} for all $n$, if $n > N$ then $A - \epsilon < a_n < A + \epsilon$ \textbf{(partial proof end)} \end{tabular}}                                                                                     & \multirow{6}{*}{$\square$}                              \\
                                                                                                                                                                                                                                                                                                                                                                                                                                                                                                                 &                                                         \\
                                                                                                                                                                                                                                                                                                                                                                                                                                                                                                                 &                                                         \\
                                                                                                                                                                                                                                                                                                                                                                                                                                                                                                                 &                                                         \\
                                                                                                                                                                                                                                                                                                                                                                                                                                                                                                                 &                                                         \\
                                                                                                                                                                                                                                                                                                                                                                                                                                                                                                                 &                                                         \\ \hline
\textbf{-} \textbf{Remove all the premises added above}                                                                                                                                                                                                                                                                                                                                                                                                                                                          & \multirow{6}{*}{$A = \lim\limits_{n \to +\infty}{a_n}$} \\
\multirow{5}{*}{\begin{tabular}[c]{@{}l@{}}\textbf{+} for all $\epsilon > 0,$ there exists $N$ such that $a_N > A - \epsilon$\\ \textbf{+} for all $\epsilon > 0$, there exists $N$ such that for all $n$, if $n > N$ then $a_n > a_N$\\ \textbf{+} for all $n$, there exists $N$ such that if $n > N$ then $a_n < A$\\ \textbf{+} for all $n$, there exists $N$ such that if $n > N$ then $A - \epsilon < a_n < A + \epsilon$\end{tabular}} &                                                         \\
                                                                                                                                                                                                                                                                                                                                                                                                                                                                                                                 &                                                         \\
                                                                                                                                                                                                                                                                                                                                                                                                                                                                                                                 &                                                         \\
                                                                                                                                                                                                                                                                                                                                                                                                                                                                                                                 &                                                         \\
                                                                                                                                                                                                                                                                                                                                                                                                                                                                                                                 &                                                         \\ \hline
\end{tabular}
\end{center}
\caption{Transformation of proof goal between lines 4-7 in Fig.\ref{ex1}. The upper part of the table shows the proof goal before reaching line 4 of the proof. The middle part of the table shows the deduction of intermediate results during the partial proof. The lower part of the table shows the proof goal after terminating the partial proof in line 7 of the proof.}
\label{tab:partial}
\end{table}

\section{The Workflow of ProofGrader} \label{4}

\subsection{Overall Architecture} \label{4.1}
\begin{figure}
\centering
\includegraphics[width=16cm]{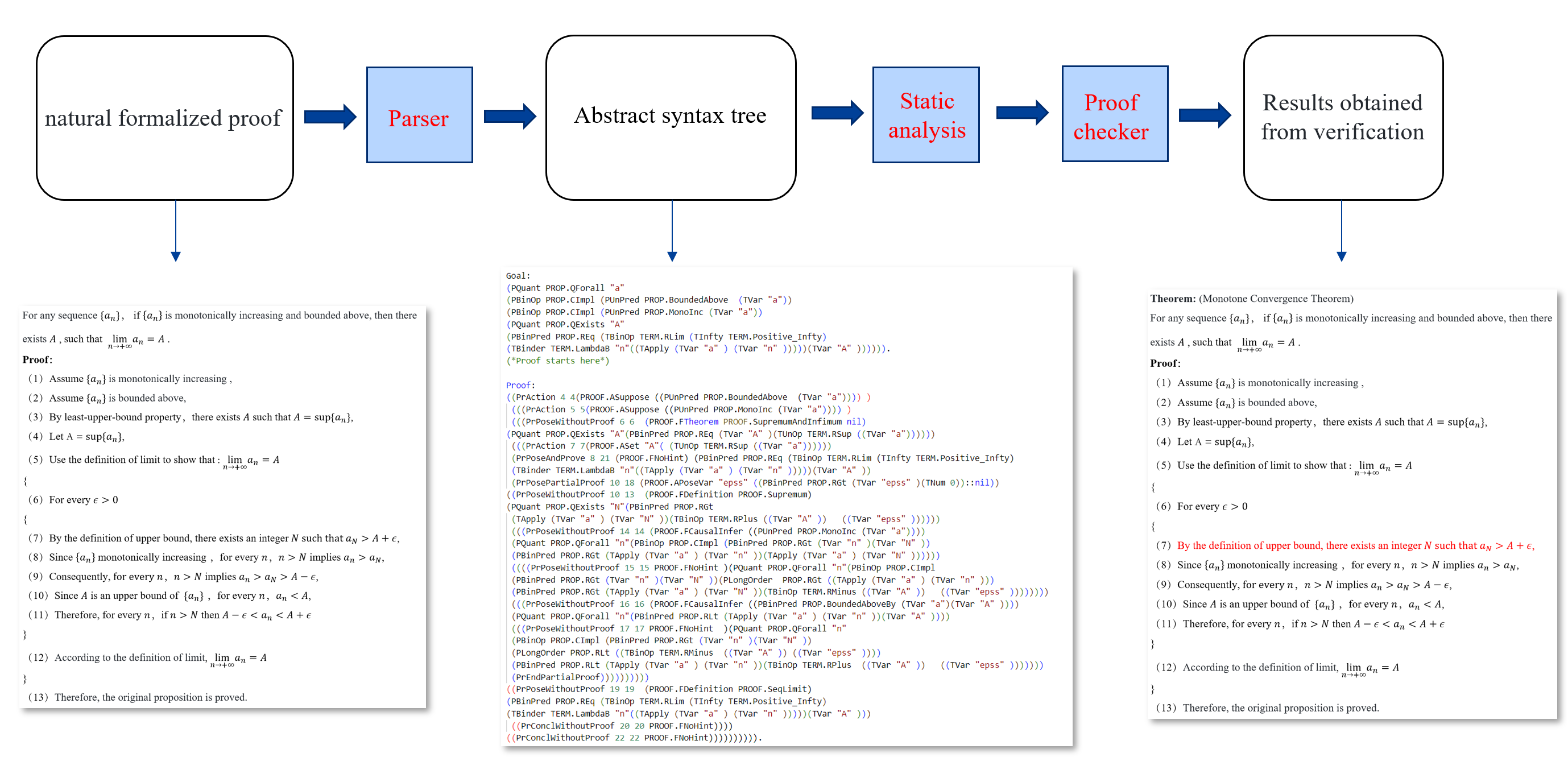}
\caption{Overall workflow of ProofGrader.} \label{arch1}
\end{figure}

In this subsection, we will primarily introduce the overall working framework of ProofGrader. Fig.\ref{arch1} illustrates how the natural formal proof is processed step by step in our proof checking system to obtain the final result. A larger figure can be found in Appendix \ref{appB} for greater clarity.

The modules described in the squares of the diagram are the main components of ProofGrader, which include the parser, static analyser, and proof checker. These three parts will be discussed in detail in the following. The other process boxes represent different forms of mathematical proof during the whole workflow. We utilize a parser to convert the natural formalized proof into an abstract syntax tree, and then perform static analysis on the abstract syntax tree to eliminate ambiguities and add omitted steps. The checker takes in the proof goal generated by the static analyser along with the complete abstract syntax tree, and checks the proof step by step by predefined rules. Finally, it confirms whether the proof goal has been proven and generates the final result.

\subsection{Proof Parsing} \label{4.2}
Proof parsing is the first step of the entire workflow of ProofGrader. Since the design of our proof language also takes a large part into account readability factors, it may not be the most convenient for the subsequent checking process. Therefore, we will first convert the natural formal proof into an abstract syntax tree described in Section \ref{2}. In our implementation, the lexer and parser are realized by flex and bison \cite{donnelly1988bison,levine2009flex}.

The parser simply performs a plain translation based on the grammar rules, without making any further modification on the proof structure or the formulation of proposition. These are the subjects of the next section where we will discuss the static analysis on the proof.

\subsection{Static Analysis} \label{4.3}

As discussed in Section \ref{intro}, natural language proofs exhibit complexities such as context-dependent semantics and overloading of notation. Given the close resemblance of our proof language to natural language and the direct translation performed by the parser, these properties will be carried into the natural formal proof, and then the abstract syntax tree. Performing static analysis on the abstract syntax tree allows us to eliminate these problems by reorganizing the proof into a more rigorous form, ready to be checked by the proof checker.

The static analyser tackles each kind of problem separately with ad hoc method. For one problematic proof step, proposition or expression, the task is to choose the right semantic between all the possible interpretations. The static analyser works by first inferring how the current proof step transforms the proof goal, based on the previous and subsequent proof goals. It then selects the semantic corresponding to this transformation, by elaborating the proof step, proposition or expression into a proper form. Judging from the problems we are currently solving by static analysis, most of them are accompanied by the appearance of free variables. So it is often the case to perform a lexical scope analysis.

\paragraph{Handling Context-dependent Semantics.} The context-dependent semantics we currently address are stated in the descriptions of Fig.\ref{ex1}. In this example, the static analyser infers the correct semantic from the context, as shown in the inference steps below. As a reminder, the three possible interpretations of ``there exists $A$ such that $A = \sup \{a_n\}$'' are respectively: (1) a value $\sup \{a_n\}$ has been found for an existential quantifier in the conclusion to be proved, (2) an intermediate result $\exists A = \sup \{a_n\}$ is stated, where $A$ is qualified by a quantifier, (3) a variable $A$ is given the value $\sup \{a_n\}$ after proving the existence of this supremum.
\begin{center}
    
\begin{tikzcd}
\boxed{\text{interpretation (1)}} \rar & \boxed{\text{proof goal beginning with } \exists} \rar{\text{analyser}} & \text{False} \\
\boxed{\text{interpretation (2)}} \rar & \boxed{A \text{ unbounded thereafter}} \rar{\text{analyser}} & \text{False} \\
\boxed{\text{interpretation (3)}} \rar & \boxed{A \text{ bounded thereafter}} \rar{\text{analyser}} & \text{True}
\end{tikzcd}

\end{center}

\noindent The three potential semantic interpretations appear syntactically identical in the proof. Therefore, the static analyser takes the responsibility for reflecting the result of the above analysis on the abstract syntax tree through modification. This is where the proof action \texttt{AExistVar} comes into play. Placed immediately after a proposition starting with an existential quantifier, it indicates the instantiation of the variable mentioned by the quantifier, thus bringing about the semantics corresponding to case (3), namely binding variable $A$ of the value.

Algorithm \ref{algoHCDS} illustrates the procedure of handling context-dependent semantics in the cases of \textbf{PoseWithoutProof} and \textbf{PoseAndProve}, the processing of other cases follows a similar approach. 


\begin{breakablealgorithm}
\caption{HCDS: Handling Context-dependent Semantics}\label{algoHCDS}
\renewcommand{\algorithmicrequire}{\textbf{Input:}}
	\renewcommand{\algorithmicensure}{\textbf{Output:}}
\begin{algorithmic}
  \REQUIRE the original proof $Pr$ before static analysis
  \ENSURE the modified proof with ontext-dependent Semantics eliminated
\STATE \textbf{When} $Pr$ = PoseWithoutProof $fwd$ $P0$ $Pr0$

\STATE  \hspace{35pt}$P0$ = PQuant exists $``x"$ $P1$

\STATE  \hspace{35pt}$``x"$ occur freely in $Pr0$

\STATE  \textbf{Then} HCDS($Pr$) = PoseWithoutProof $fwd$ $P0$ (ProofAction (AExistVar $``x"$)  HCDS($Pr0$))
\newline

\STATE \textbf{When} $Pr$ = PoseAndProve $fwd$ $P0$ $Pr0$ $Pr1$

\STATE  \hspace{35pt}$P0$ = PQuant exists $``x"$ $P1$

\STATE  \hspace{35pt}$``x"$ occur freely in $Pr1$

\STATE  \textbf{Then} HCDS($Pr$) = PoseAndProve $fwd$ $P0$  (HCDS($Pr0$) (ProofAction (AExistVar $``x"$) HCDS($Pr1$)) 

...
\end{algorithmic}
\end{breakablealgorithm}

\paragraph{Handling Overloading of Notation.} If there is only one possible interpretation for notation overloading, there is no need for analysis and the static analyser simply restores its rigorous form. So far, the cases we have encountered with multiple possible interpretations for notation overloading all involve the use of formal variables. In such cases, the static analyser examines the appearance of free variables. As an example, the $\{a_n\}$ in line 2 of Fig.\ref{ex1} admits two possible interpretations: (1) the singleton $\{a_n\}$, (2) or a set containing all elements of the sequence $a$. The correct interpretation depends on whether the variable $n$ is bound to a value in the current proof goal. Similarly, if $x$ is identified as a free variable in the proof goal, $f(x)$ will be transformed into either $f$ or $\lambda x.f(x)$, rather than the value of $x$ applied to the function $f$.

Algorithm \ref{algoHON} illustrates the procedure for handling the notation overloading of $f(x)$ in the cases of \textbf{PoseWithoutProof} and \textbf{ProofAction} $(\textbf{ASet} \; x \; t)$, the processing of other cases follows a similar approach.

\begin{breakablealgorithm}
\caption{HON: Handling Overloading of Notation}\label{algoHON}
\renewcommand{\algorithmicrequire}{\textbf{Input:}}
	\renewcommand{\algorithmicensure}{\textbf{Output:}}
\begin{algorithmic}
  \REQUIRE the original proof $pr$ before static analysis

  \hspace{16pt} the list of binded variables $binded\_varlist$ from the beginning to the current point of the proof 
  \ENSURE the modified proof with overloading of notation eliminated
  \STATE \textbf{When} $Pr$ = PoseWithoutProof $fwd$ $P0$ $Pr0$

\STATE  \hspace{35pt}$P0$ = Eq (TApply ($``f"$) ($``x"$)) $t1$

\STATE  \hspace{35pt}if $``x"$ is not in $binded\_varlist$

\STATE  \textbf{Then} HON($Pr$, $binded\_varlist$) = PoseWithoutProof $fwd$ (Eq ($f$) (\textbf{TBinder Lambda} $``x"$ $t1$)) HON($Pr0$, $binded\_varlist$)
\newline

 \STATE \textbf{When} $pr$ = ProofAction (ASet $``x"$ $t$) $Pr0$ 
 \STATE  \textbf{Then} HON($Pr$, $binded\_varlist$) = ProofAction (ASet $``x"$ $t$) HON($Pr0$, $x::binded\_varlist$)
 
...

\end{algorithmic}
\end{breakablealgorithm}

\subsection{Proof Checking} \label{4.4}

\begin{sidewaysfigure}[p] 
    \centering
    \includegraphics[width=\columnwidth]{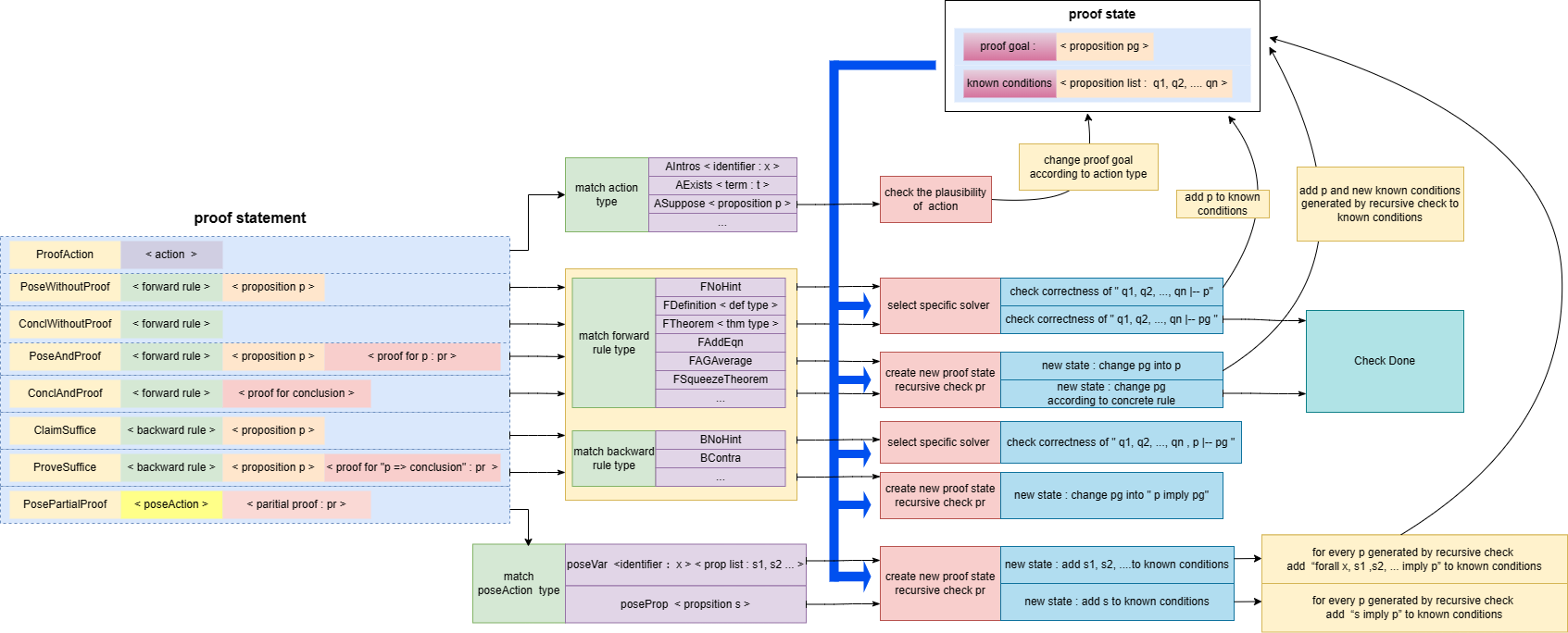} 
    \caption{Process of proof checking.}
    \label{fig:checker}
\end{sidewaysfigure}

Proof checking is the final step of the entire workflow of ProofGrader. The checker takes the proof and the proof goal elaborated by the static analyser as input. For each proof step, the checker takes the current proof goal and checks the step according to the formal semantics presented in Section \ref{3}, it computes the subsequent proof goal along with a boolean value indicating whether the step is accepted. By iteratively repeating this process, the proof checker finally generates a list of boolean values indicating the correctness of each proof step. In order to help with proposition checking, we also develop several solvers and a solver manager system. We will introduce them in Section \ref{5}.
The entire proof checking process is detailed in Fig.\ref{fig:checker}.

\section{Solver Manager} \label{5}
The solver manager is an important component of the proof checker, and therefore, we will introduce it separately here. It is primarily used to verify the correctness of mathematical propositions under certain conditions. When it comes to proving and simplifying mathematical expressions, we combine different solvers to check if the current proposition can be derived from known results using specific rules.

Each solver corresponds to a deduction rule based on specific mathematical concepts, and implements the corresponding algorithm internally in the proof checker. Different solvers are responsible for handling different types of mathematical expressions. For example, when dealing with expressions involving algebraic operators, ProofGrader utilizes algebra-related solvers for simplification. Similarly, solvers for trigonometric functions, exponential and logarithm, sequential limit, and other mathematical concepts are employed to handle corresponding cases.

When users write proofs, they must adhere to the steps supported by the solvers. In Section \ref{intro}, it was noted that AI-generated formal proofs sometimes use solvers like Z3\cite{de2008z3} to check propositions. However, Z3 possesses strong reasoning capabilities and can sometimes prove the correctness of conclusions even if the original deduction process contains error. On the other hand, our solver does not engage in excessively powerful reasoning. It only supports relatively obvious deduction methods in proofs and does not allow excessive omission of steps. This approach aims to reflect the correctness of the original proof more accurately. Additionally, since we can provide explicit deduction rules supported by each solver, users can clearly know which steps can be omitted and which steps cannot be omitted in their proof process. 

\subsection{Structural Design of the Solver} \label{5.1}
Before presenting the design of the solver manager, it is necessary to first define the structure of the solver, as these two are closely related. Each solver consists of the following four components:
\begin{itemize}
    \item \emph{solve(s,p)} : the function used to verify the correctness of a proposition takes two parameters: the current proof goal and the proposition itself. It can either return false when the proposition does not pass the solver, or a list of propositions, indicating that after being processed by the current solver, the correctness of the original proposition can be determined by individually checking the propositions in the list. This design allows us to combine multiple solvers to simplify a proposition. The list can contain only one proposition, which means the solver simplifies the original proposition and passes it to the next solver. An empty list indicates that the current proposition is accepted by the solver.
    \item \emph{fee} : the cost of this solver. While we want to combine multiple solvers, we cannot endlessly repeat using various solvers, as it may result in non-termination. Therefore, we need to set an upper limit on the total cost within the solver manager and define the cost of each solver.
    \item \emph{default-priority} : the default priority of solvers. When dealing with mathematical propositions that do not exhibit any prominent characteristics, solvers can be selected based on their default priority levels.
    \item \emph{priority(p)} : the function used to compute the dynamic priority. Given the current proposition as input, this function returns the dynamic priority of the solver. For example, when handling a mathematical proposition without limits, solvers related to sequential limit may have a low priority or even be unavailable. However, when dealing with limit-related propositions, these solvers would have a high priority. Therefore, it is necessary to dynamically adjust the priority of solvers based on the specific proposition. This prepares us for the subsequent development of the solver manager.
    
\end{itemize}

\subsection{The Design of Solver Manager} \label{5.2}

The solver manager achieves dynamic scheduling of various solvers by combining dynamic priorities. First, we categorize the existing solvers into two main types. The first type is the general solvers, which are applicable to various forms of propositions. For example, they can determine if both sides of an equation are identical or simplify expressions using polynomial rules. The second type is conditional solvers, which are designed to handle propositions that meet specific forms, or involve a specific mathematical object, field of knowledge, etc.

When examining the current proposition, the solver manager needs to provide a list of available solvers related to that proposition. The algorithm proceeds as follows: for each conditional solver, the dynamic priority of the solver is computed for the current proposition. Then, all conditional solvers with a non-zero dynamic priority are included in the available solver list. Finally, all general solvers are appended to the list, forming the final list of available solvers. This is the functionality implemented by the $usablelist$ function in Algorithm \ref{algo1}.

\begin{algorithm}
\caption{Implementation of the solver manager} \label{algo1}
\renewcommand{\algorithmicrequire}{\textbf{Input:}}
	\renewcommand{\algorithmicensure}{\textbf{Output:}}
\begin{algorithmic}[1]
  \REQUIRE  $fee$: limit on the total cost; 
  \hspace{20pt}$ls$ : list of solvers;
  
  \hspace{20pt}$pg$ : the proof goal;
  \hspace{50pt}$p$ : proposition to be verified;
  \ENSURE the correctness of proposition p;
  
\STATE \textbf{Procedure} SolverManager ($fee$, $ls$, $pg$, $p$)
 \STATE  \textbf{if} $fee > 0$ \textbf{then}
 \STATE \hspace{10pt} $uselist$ = usablelist($ls$, $p$);
 \STATE \hspace{10pt}  \textbf{for} every solver $s$ in $uselist$ \textbf{do}
 \STATE  \hspace{20pt}  $listProp$ = s.solve($pg$, $p$);
 \STATE  \hspace{20pt} \textbf{if} $listProp$ is empty \textbf{then} return true;
 \STATE  \hspace{20pt} \textbf{else if} $listProp$ is false
 \textbf{then} return false;
 \STATE  \hspace{20pt} \textbf{else} $flag$ = false;
 \STATE \hspace{30pt} \textbf{for} every proposition $p'$ in $listProp$
 \STATE \hspace{40pt} \textbf{if} $p'$ == $p$ \textbf{then} break;
 \STATE \hspace{40pt} \textbf{else if} SolverManager($fee - s.fee$, $ls$, $pg$, $p'$) is false;
 \STATE \hspace{50pt} $flag$ = true;
 \STATE \hspace{50pt} break;
 \STATE \hspace{30pt} \textbf{if} $flag$ is true \textbf{then} return true;
 \STATE \hspace{10pt} return false;
 \STATE \textbf{else} return false;

\end{algorithmic}
\end{algorithm}

\noindent As shown in Algorithm \ref{algo1} above, the final solver manager takes in four parameters and returns the correctness of the proposition. The manager first checks the remaining cost. If it is greater than $0$, it selects a subset of solvers from $ls$ to form an available solver list $L$ by calculating the dynamic priority of each solver based on the form of proposition $p$. Let $L$ be $(s1,s2,s3,...,sn)$. Then, it selects $s1$ and uses $s1$'s solver function to check proposition $p$. If the result is an empty list or false, it returns the proposition as true or false. If it returns a list of propositions $listProp$, it means the original proposition has been simplified and decomposed to be passed to other solvers for resolution. Before passing it to other solvers, we need to check if $listProp$ contains an identical proposition to the original proposition $p$. If it exists, it means that $s1$ did not make any effective contribution to the validation of proposition $p$, so this branch will be pruned, and the solver manager will continue to call $s2$ to handle proposition $p$. Otherwise, the solver manager recursively validate each proposition in $listProp$. If all validation results are true, the original proposition $p$ is true. If any sub-proposition is false, this branch will also be pruned, and the solver manager will continue to call $s3$ to recheck proposition $p$, and so on until the final validation is completed.

\textbf{Configurability adapted to educational scenarios.} Since the initial list of solvers that the solver manager can use is configurable, we can conveniently control the deduction steps supported by the checker. When applied in an educational setting, we can set the solver list based on the progress of the curriculum, so that students can only use the methods learned in the current chapter to complete the proofs, and other methods will not pass the checker. For instance, if students are expected to evaluate limits by the definition, it is necessary to disable solvers like those for L'Hôpital's rule. This management design can also be used to regulate which theorems can be used in the proof checker.

\section{Evaluation} \label{evaluation}

In this section, we give an evaluation of our proof checking system as well as our proof language. We first run our system on a set of sample mathematical proofs, covering the topic of arithmetic, trigonometric functions, exponential and logarithm, inequality, derivative, sequential limit and function continuity. The source code of our system and the dataset are available at: \href{https://github.com/Laplace-Demon/ProofGrader}{https://github.com/Laplace-Demon/ProofGrader}. We then compare the features of our proof language with other proof languages and proof checking tools.

\subsection{Performance}

We test our system on a dataset of 52 mathematical proofs, Table \ref{tab:eval} shows the average time and memory overhead of the proof parser and the checker on each of the topics, along with the average file size of the proof. We can observe that our proof parser and checker perform their task in a reasonable amount of time.

\begin{table}[ht!]
\begin{center}
\begin{tabular}{|ccc|cc|cc|}
\hline
\multicolumn{3}{|c|}{Examples}                                                                      & \multicolumn{2}{c|}{Parser}                  & \multicolumn{2}{c|}{Checker}                 \\ \hline
\multicolumn{1}{|c|}{Topic}                       & \multicolumn{1}{c|}{Number} & File size (bytes) & \multicolumn{1}{c|}{Time (ms)} & Memory (kb) & \multicolumn{1}{c|}{Time (ms)} & Memory (kb) \\ \hline
\multicolumn{1}{|c|}{arithmetic}                  & \multicolumn{1}{c|}{6}      & 271               & \multicolumn{1}{c|}{34}        & 3584        & \multicolumn{1}{c|}{477}       & 3456        \\ \hline
\multicolumn{1}{|c|}{trigonometric functions}     & \multicolumn{1}{c|}{8}      & 599               & \multicolumn{1}{c|}{53}        & 3712        & \multicolumn{1}{c|}{1409}      & 3456        \\ \hline
\multicolumn{1}{|c|}{exponential and logarithm}   & \multicolumn{1}{c|}{3}      & 371               & \multicolumn{1}{c|}{40}        & 3840        & \multicolumn{1}{c|}{506}       & 3456        \\ \hline
\multicolumn{1}{|c|}{inequality}                  & \multicolumn{1}{c|}{10}     & 395               & \multicolumn{1}{c|}{58}        & 3712        & \multicolumn{1}{c|}{528}       & 3456        \\ \hline
\multicolumn{1}{|c|}{derivative}                  & \multicolumn{1}{c|}{3}      & 385               & \multicolumn{1}{c|}{75}        & 3328        & \multicolumn{1}{c|}{760}       & 3328        \\ \hline
\multicolumn{1}{|c|}{sequential limit evaluation} & \multicolumn{1}{c|}{10}     & 550               & \multicolumn{1}{c|}{58}        & 3840        & \multicolumn{1}{c|}{701}       & 3328        \\ \hline
\multicolumn{1}{|c|}{sequential limit proof}      & \multicolumn{1}{c|}{10}     & 1027              & \multicolumn{1}{c|}{78}        & 3968        & \multicolumn{1}{c|}{670}       & 3328        \\ \hline
\multicolumn{1}{|c|}{function continuity proof}   & \multicolumn{1}{c|}{2}      & 923               & \multicolumn{1}{c|}{68}        & 3712        & \multicolumn{1}{c|}{7758}      & 3456        \\ \hline
\end{tabular}
\end{center}
\caption{Runtime and memory usage of the parser and checker on different topics.}\label{tab:eval}
\end{table}

\subsection{Comparison of Features}\label{Comparison of Features}

In the table\ref{tab:feature} below, we compare the features of our proof language with some other proof languages and proof checking tools. The term ``transformation chain'' refers to the ability to handle and perform automated reasoning on a series of consecutive transformation derivations, such as: $$\lim\limits_{x \to 0}{\frac{\sqrt[3]{x+1}-1}{\sqrt{x+1}-1} }= \lim\limits_{x \to 0}{\frac{x}{(\sqrt[3]{x+1})^{2} + \sqrt[3]{x+1} + 1} \cdot \frac{\sqrt{x+1}+1}{x}} = \frac{2}{3}$$

\noindent Our proof checker aims to provide automated deduction capabilities that align with human judgment, while excessive step omissions are not allowed.

In terms of readability, our natural formal proof can be read with the same effort of reading a natural language mathematical proof. Even the abstract syntax tree generated by the parser remains readable since the components bear a mnemonic name. 

In terms of static analysis, most theorem provers like Coq and Isabelle primarily support propositional-level static analysis. They utilize type inference to fill in missing information in user-written proofs. However, they do not perform static analysis on the entire proof.
Regarding partial proofs, while it is sometimes possible to reorganize a proof to avoid partial proofs, we aim to preserve the characteristics of natural language proofs. Users should be able to write mathematical proofs without additional effort.

The work of Waterproof\cite{wemmenhove2022waterproof} aims to help students understand mathematical proofs. Therefore, its language readability falls between natural language and theorem provers. The proof checking functionality of Waterproof relies on Coq implementation, so it inherits some limitations of Coq as well. Diproche\cite{carl2020diproche} utilizes natural language fragments as its input language. Due to its lack of support for static analysis and partial proofs, it imposes stricter requirements on the types and structure of proofs being written. Lurch\cite{carter2014using} is limited to checking proofs in propositional logic and naive set theory. It does not have automatic reasoning capabilities and relies on users to provide detailed proofs for checking.

\begin{table}[ht!]
\begin{center}
\begin{tabular}{|c|c|c|c|c|c|c|}
\hline
Feature & ProofGrader & WaterProof & Diproche &  Lurch & Isabelle\cite{paulson1994isabelle} & Coq\cite{barras1999coq}\\ \hline
Natural language fragment  & $\checkmark$  &$\times$   & $\checkmark$ & $\checkmark$   &$\times$ &$\times$\\ \hline
Static analysis  & proof-level   &$\times$  &$\times$  &$\times$  & prop-level & prop-level \\ \hline
Partial proof  & $\checkmark$  &$\times$ &$\times$  &$\times$  &$\times$ &$\times$ \\ \hline
Transformation chains & $\checkmark$  &$\times$ & $\checkmark$  &$\times$  &$\checkmark$ &$\times$ \\ \hline

\end{tabular}
\end{center}
\caption{Comparison of features of different mathematical proof checking tools.}\label{tab:feature}
\end{table}

\section{Related Work} \label{related}

\paragraph{Proof assistants and other proof checkers.} Recent decades have seen the emergence of various proof languages. A well-known work is Ltac\cite{David2000Ltac} for the theorem prover Coq, which provides convenience for constructing proofs and facilitates better proof automation. Another famous one is that of Isar\cite{Wenzel1999IsarA} for the system Isabelle. One objective of Isar is to provide a more human-readable proof language than before. Despite their efforts to improve readability, the learning curve for using these tools remains high. Wemmenhove et al. developed an educational software called Waterproof\cite{wemmenhove2022waterproof} based on this to assist students in practicing proofs. Since they still choose to rely on the tactic library extended with the Ltac2 tactic language, Waterproof has not actually gained stronger expressive power than these tactic-based proof language. Additionally, the proof assistant Agda\cite{bove2009brief} also employs some proof notations that resemble natural language, such as using equation chain instead of the ``rewrite'' tactic used in Coq. However, despite these advancements, they still struggle to effectively support the structure of partial proofs. While they can perform static analysis at the propositional level, such as type inference, they are unable to perform static analysis on the entire proof. In addition, tools mentioned in Section\ref{Comparison of Features}, such as Diproche and Lurch, although they support controlled natural language input, have stricter requirements on the notation and structure of proofs being written.

\paragraph{Machine learning for formalization.} Machine learning techniques have been used in the formalization of informal proofs. The work of Yuhuai Wu et al.\cite{wu2022autoformalization} based on large language models can correctly transform 25.3\% of the solution for mathematical competition problems in MATH dataset\cite{hendrycks2021measuring} into formal specifications in Isabelle/HOL. In the work of Jiang et al.\cite{jiang2022draft}, they introduce Draft, Sketch, and Prove (DSP), a method that maps informal proofs to formal proof sketches, The accuracy was improved to 39.3\% on the same dataset\cite{hendrycks2021measuring}. And the automatic theorem prover implemented by GPT-f\cite{polu2020generative} achieved a completion rate of 56.22\% on their test set. The purpose of formalizing proofs with AI-based methods is to ensure that the proven proposition is correct, especially when proving previously unproven propositions in mathematical research. Since the target language of the transformation is often tactic-based proof languages that differ greatly from natural language, these works mainly use the original proof to guide theorem provers to complete the proof, rather than truly transforming the original proof into a formalized proof. In this case, because the language we designed combines formal rigor with similarity to natural language,  if AI automatic translation is set to use our language as the target, it will complement our work well and lead to better results.


\section{Conclusion} \label{Conclusion}
In this paper, we present our design of a natural-formalized proof language and the implementation of a mathematical proof checker. Compared to existing tactic-based proof language, our proof language has more expressive power thanks to the incorporation of partial proof. To cope with the characteristics of natural language proof, such as context-dependent semantics and overloading of notation, tactic language provides a fine-grained but cumbersome formal specification in the hope that someone can correctly reproduce the proof, while our proof language can automatically fix these problems through static analysis. All these factors result in better readability and an easier formalization process.

Regarding the process of proof checking, we implement a solver manager responsible for managing various automated proof checking strategies. It selects the most suitable strategies based on the form and contextual environment of the proposition to assess its correctness. Building upon this, the proof checker takes the proof and the proof goal provided by the static analyser, applies the appropriate checking methods, updates the proof goal iteratively, and ultimately completes the checking process.

\clearpage
\bibliographystyle{splncs04}
\bibliography{reference}

\clearpage
\appendix

\section{ Key grammar of natural formalized proof language} \label{AA}

\begin{grammar}

<proof> ::= <proof statement>
\alt <proof statement> `,' <proof>
\alt `The following proves' <math proposition> `{' <proof> `}' <proof>
\alt `It suffices to prove' <math proposition> <proof>
\alt <proof_action_statement> `{' <proof> `}' <proof>

<proof statement> ::= `which proves the proposition.'
\alt `Use' <definition> `to prove'
\alt `Use' <theorem> `to prove'
\alt <since_clause> `then' <math proposition>
\alt <proof_action_statement> 
\alt ...

<since_clause> ::= `Obviously'
\alt `Similarly'
\alt `Since' <knowledge>
\alt `Because' <math proposition>
\alt `By' <prop_action_statement> 
\alt ...

<prop_action_statement> ::= adding <list proposition name>
\alt <prop_transformation> `on the both sides of the equality'
\alt ...

<prop_transformation> ::= `adding' <math expression>
\alt `squaring'
\alt `taking the logarithm'
\alt `taking the derivative of' <variable name>
\alt ...

<proof_action_statement> ::= `Let' <variable name>
\alt  `Let' <variable name> <math proposition>
\alt `There exists' <math expression>
\alt `There exists' <variable name> `such that' <math proposition>
\alt `Suppose' <math proposition>
\alt `Suppose' <variable name> `such that' <math proposition>
\alt `Set' <variable name> `=' <math expression>
\alt ...

<knowledge> ::= <theorem>
\alt <definition>
\alt <property>

\end{grammar}
Since natural language offers various ways to express the same meaning, we have chosen a subset of these expressions as keywords for the grammar of natural formalized proof. In the above grammar, we have chosen only one of the possible options as example. Furthermore, due to the complexity of mathematical expressions and propositions involving various mathematical concepts, we do not elaborate on them in detail here.
\clearpage
\section{Grammar of \texttt{term} and \texttt{prop}} \label{BB}

\begin{grammar}

<term> ::= `TNum' <number>
\alt `TInfty' <infinity>
\alt `TConst' <const> 
\alt `TUnOp' <unaryTermOperator> <term>
\alt `TBinOp' <binaryTermOperator> <term> <term>
\alt `TApply' <term> <term>
\alt `TBinder' <binder> <identifier> <term>
\alt `TVar' <identifier>
\alt `TInterval' <intervalType> <term> <term>
\alt `TSet' \{\textit{string}\}$^*$ <term> <prop>

<prop> ::= `PUnPred' <unaryPredicate> <term>
\alt `PBinPred' <binaryPredicate> <term> <term>
\alt `PCBinPred' <binaryPredicate> <term> <term> <propContext>
\alt `PLongOrder' <order> <term> <prop>
\alt `PUnOp' <unaryPropOperator> <prop>
\alt `PBinOp' <binaryPropOperator> <prop> <prop>
\alt `PQuant' <quantifier> <identifier> <prop>

<binder> ::= `SeqLimitB'
\alt `LambdaB'

\end{grammar}

\clearpage


\begin{sidewaysfigure}[p] 
    \section{The overall workflow of ProofGrader}\label{appB}
    \centering
    \includegraphics[width=\columnwidth]{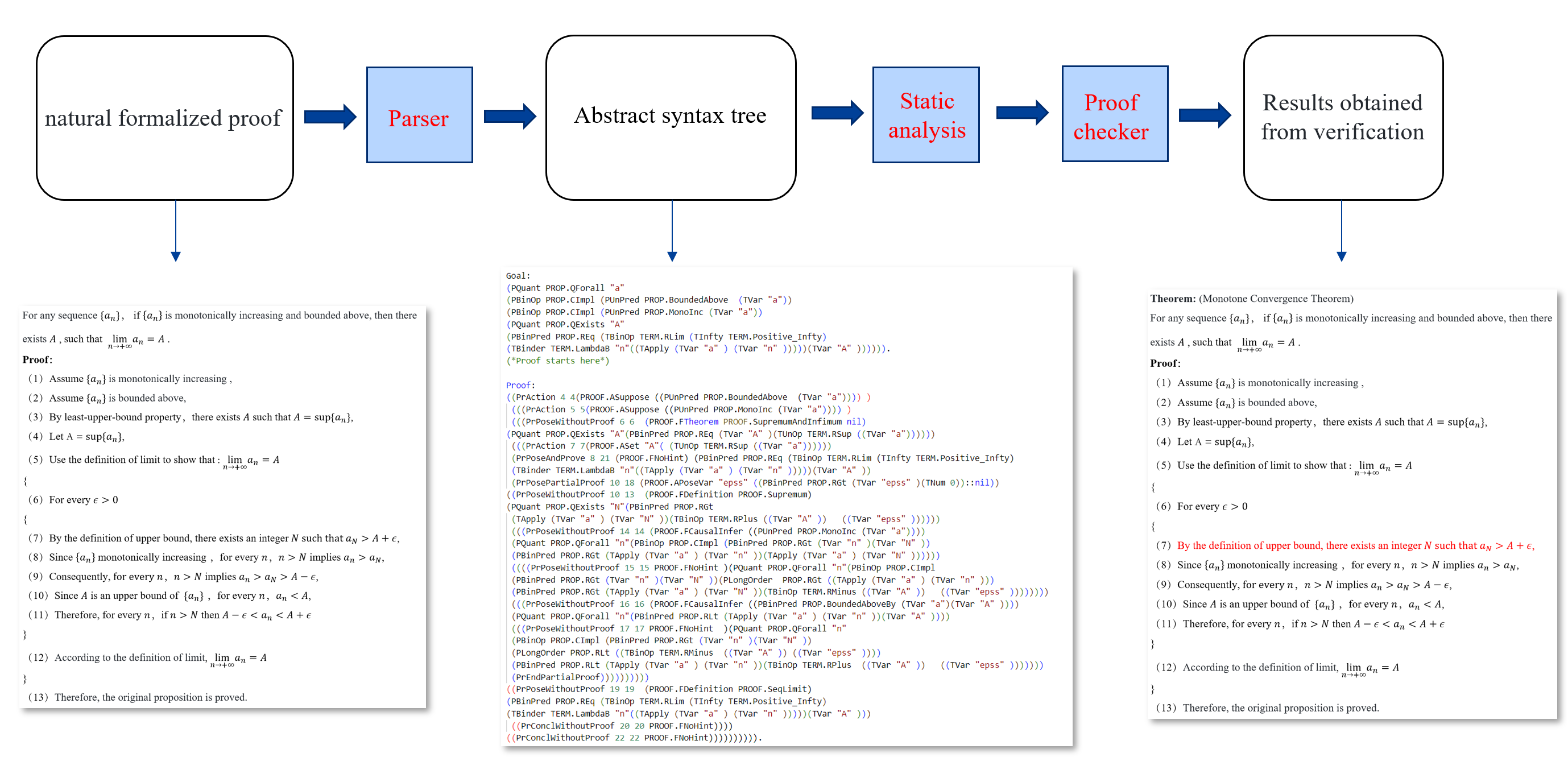} 
    \label{fig:fig1}
    \caption{Overall workflow of ProofGrader.}
\end{sidewaysfigure}


%
%
%

%



\end{document}